\documentclass[prb,aps,12pt,a4paper,showpacs]{revtex4}

\usepackage{epsfig,color}

\def\RR#1{#1}

\begin{document}

\title{Level compressibility in a critical random matrix ensemble: \\
          The second virial coefficient
      }

\author{Vladimir E. Kravtsov}
\email{kravtsov@ictp.it}
\affiliation{ The Abdus Salam ICTP, Strada Costiera 11, 34100, Trieste, Italy \\
                    Landau Institute for Theoretical Physics, 2 Kosygina st.,
                    117940 Moscow, Russia
                  }

\author{Oleg Yevtushenko}
\email{bom@ictp.it}
\affiliation{The Abdus Salam ICTP, Strada Costiera 11, 34100, Trieste, Italy}

\author{Emilio Cuevas}
\email{ecr@um.es}
\affiliation{Departamento de F{\'\i}sica, Universidad de Murcia, E-30071 Murcia, Spain}

\date{\today}

\begin{abstract}
We study spectral statistics of a Gaussian unitary critical ensemble 
of almost diagonal Hermitian random matrices with off-diagonal
entries $\langle|H_{ij}|^{2} \rangle \sim b^{2}\, |i-j|^{-2}$
small compared to diagonal ones $\langle|H_{ii}|^{2} \rangle \sim
1$. Using the recently suggested method of  {\it virial expansion} in
the number of interacting energy levels (J.Phys.A {\bf 36},8265
(2003)), we calculate a coefficient $\propto b^{2}\ll 1$ in the
level compressibility $\chi(b)$. We demonstrate that only the
leading terms in $\chi(b)$ coincide for this model and for an
exactly solvable model suggested by Moshe, Neuberger and Shapiro
(Phys.Rev.Lett. {\bf 73}, 1497 (1994)), the sub-leading terms
$\sim b^{2}$ being different. Numerical data confirms our
analytical calculation.
\end{abstract}

{}

\maketitle

\section{Introduction}

\subsection{Critical power-law banded random matrices and
exactly solvable models}

Recently, there was an increasing interest to unconventional
random matrix theories (RMTs) that interpolate between the
Wigner-Dyson RMT and banded RM with the (almost) Poissonian level
statistics. One of these models is the {\it power law banded
random matrix} theory \cite{MF,KM,ME} for which the variance of
the off-diagonal elements takes the form:
\begin{equation}\label{VarPLperiodic}
  {\rm PLBRM:} \quad
  \langle | V_{ij}|^{2} \rangle = { 1 \over 2 } \frac{1}{ 1 +
                              \left(
                        \frac{N}{\pi \, b} \sin \left( \frac{\pi}{N} |i-j| \right)
                              \right)^{2\alpha}
                                          } \, .
\end{equation}
Here $ \, N \, $ is the matrix size; $ \, \alpha \mbox{ and } b \,
$ are two parameters which control statistical properties of
PLBRM. The variance (\ref{VarPLperiodic}) is nearly constant
inside the band $ \, |i-j| < b $, and decreases as a power-law
function $\langle |V_{ij}|^{2}\rangle \sim 1/|i- j|^{-2 \alpha}$
for $ \, |i-j|> b $.  The case $\alpha >1$ corresponds to the
power-law localization which can be found in certain periodically
driven quantum-mechanical systems \cite{KickRotPLRM}. If $ \,
\alpha \le 1/2 \, $ the spectral statistics of PLBRM approaches
that of the Wigner-Dyson RMT. The special case $ \, \alpha=1 \, $
is relevant for description of critical systems with multifractal
eigenstates \cite{MF,KM,ME,AltLev,Levitov}, in particular for
systems at the Anderson localization-delocalization transition
point. On the other hand, it has been conjectured \cite{KrTs} that
the spectral statistics of critical PLBRM with large $ \, b \, $
can be mapped onto the Calogero-Sutherland model (CS)
[\onlinecite{CS}] at low temperature $T\sim 1/b$. According to
this mapping instead of the spectral problem of random matrices
one studies the statistics of one-dimensional fermions in a
parabolic confinement potential interacting with the inverse
square potential $(x_{i}-x_{j})^{-2}$ (for real off-diagonal
elements in PLBRM, or {\it the orthogonal ensemble }) or
non-interacting (for complex off-diagonal elements in PLBRM with
identical distribution of real and imaginary parts, or {\it the
unitary ensemble}).

However, there is an RMT for which the mapping onto the CS model
is {\it exact } \cite{Garcia}. This is the model of Moshe,
Neuberger and Shapiro (MNS) [\onlinecite{MNS}]. The probability
distribution of the Hamiltonian $ \, \hat {\cal H} \, $ in MNS is
given by $P(\hat{{\cal H}})=\int {\rm d}\hat{U} \ {\cal P}_{\hat
U} ( \hat {\cal H} ) $, where
\begin{equation}\label{MNS-def}
 {\cal P}_{\hat U} ( \hat {\cal H} ) \propto \exp \Biggl( - {\rm Tr}
            \hat{{\cal H}}^2 -
            \left( \frac{N}{ 2 \pi b } \right)^2 {\rm Tr} \Bigl( [ \hat{U},
            \hat{{\cal H}} ] [ \hat{U}, \hat{{\cal H}} ]^{\dagger} \Bigr)
            \Biggr) \, ;
\end{equation}
the matrix $ \, \hat U \, $ is either unitary ( for ${\cal H}$
from the unitary ensemble) or orthogonal (for ${\cal H}$ from the
orthogonal ensemble), and $ \, {\rm d}\hat{U} \, $ is the Haar
measure.

The connection between PLBRM and MNS is especially clear in the
unitary case \cite{KM} where the unitary matrix $\hat U=M\,
diag\{e^{i\varphi_{i}} \}\,M^{\dagger}$ can be diagonalized by a
unitary transformation. Then the variances of $ \, V_{i,j} =
\left( M^{\dagger} \hat{{\cal H}} M \right)_{i,j} \, $ in MNS at
given phases $ \, \varphi_i \, $ are:
\begin{equation}\label{VarMNS}
  {\rm MNS:} \quad
  \langle | V_{ij}|^{2} \rangle = { 1 \over 2 } \frac{1}{ 1 +
                              \left(
                        \frac{N}{\pi b}\right)^{2}\, \sin^{2} \left(
\frac{\varphi_i-\varphi_j}{2}\right)
                                          } \, .
\end{equation}
One can easily see that Eq.(\ref{VarMNS}) coincides with
Eq.(\ref{VarPLperiodic}) at $\alpha=1$ if the phases
$\varphi_{n}=2\pi n/N$ are arranged as an ordered array on a
circle. In general, the MNS model can be considered as an
extension of the PLBRM model for the case of a random arrangement
of phases $\varphi_{n}$ distributed over the circle with the joint
probability distribution $ \,  P(\{\varphi\}) $
[\onlinecite{MNS}]:
\begin{equation} \label{JPD}
   P(\{\varphi\}) \sim \prod_{i>j}
                    \frac{ \sin^2 \left( \frac{\varphi_i-\varphi_j}{2}
                       \right) }{ 1 +
                              \left(
                        \frac{N}{\pi b}\right)^{2}\, \sin^{2} \left(
   \frac{\varphi_i-\varphi_j}{2}\right)} \, .
\end{equation}
The  averaged value of an observable $ \, A(\hat{H}) $, which is invariant
under the transformation $ \, \hat{H} \to M^{\dagger} \hat{{\cal H}} M $,
can be calculated as:
\begin{equation}\label{MNSaver}
   \langle \langle A \rangle_{\hat{H}} \rangle_{\hat{U}} \equiv
       {
         \int \langle A \rangle_{\hat{H}} \, P(\{\varphi_{i} \})\, {\rm D}
\{
\varphi_i \} \over
         \int P(\{\varphi_{i}\}) \, {\rm D} \{ \varphi_i \}
       } \, .
\end{equation}
Here $\langle A \rangle_{\hat{H}}$ stands for the averaging over the
Gaussian random matrix $\hat{H}$ with entries having zero mean value and
the variance given by Eq.(\ref{VarMNS}).

The two-point correlation function, which follows from Eq.(\ref{JPD}) after the
integration over all but two phases, was calculated by Gaudin with the help of the
model of free non-interacting fermions with a linear spectrum \cite{Gaudin}:
\begin{equation} \label{PhR2}
   {\cal R}_2 ( s ) = 1 - \frac{1}{( 2 \pi b )^2 }
                   \left|
                \int_{ - \log \left( e^{2 \pi b} - 1 \right) }^{\infty}
                   \frac{ e^{ {\bf i} { \omega s \over b } } \, {\rm d} \omega }
                        { e^{\omega} + 1 }
                   \right|^2 \, , \quad
  s \equiv (\varphi_i-\varphi_j)(N/2\pi) \, .
\end{equation}
If $ \, | s | \gg b $, the correlation function is almost constant $ \, {\cal R}_2
\Bigl( | s | \gg b \Bigr)  \to 1 $. There is a repulsion between phases at a small
scale controlled by $ \, b $: $ \, {\cal R}_2 \Bigl( | s | \ll b \Bigr) \sim ( s /
b )^2 $.

\subsection{Spectral statistics of MNS and PLBRM}

The level statistics of RMT is characterized by the density of states
\begin{equation}\label{DOS-E}
  \rho(E) = \langle \, \sum_{n=1}^N \delta( E - \epsilon_n ) \, \rangle \, ,
\end{equation}
and its multi-point correlation functions.
For example, the two-level correlation function $ \, R(\omega) \, $ is defined as:
\begin{equation}
\label{R-corr}
     R(\omega) = { \langle \langle \, \rho(\omega/2) \, \rho(-\omega/2) \, \rangle
\rangle \over \langle \,
                     \rho(0) \, \rangle^{2} } \, ; \quad
     \langle \langle \, \hat{a} \, \hat{b} \, \rangle \rangle \equiv \langle \,
\hat{a} \,
\hat{b} \,
            \rangle - \langle \, \hat{a} \, \rangle \langle \, \hat{b} \, \rangle \, .
\end{equation}
The Fourier transform of $ \, R(\omega) \, $ is known as the
spectral form-factor~$ \, K(t) $:
\begin{equation}
   K(t) = \int_{-\infty}^{+\infty} e^{{\bf i}\,\omega t} \, R( \omega ) \,
{\rm d} \omega \, .
\end{equation}
We rescale time by the mean level spacing
\begin{equation}\label{Delta}
    \Delta \equiv \frac{1}{\langle \, \rho(0) \, \rangle}
\end{equation}
introducing the dimensionless time $ \, \tau = t \, \Delta $.  In the limit
of small time the spectral form-factor $ \, K( \tau \to 0 ) \, $ is linked to the other
important spectral characteristics called {\it the level compressibility} \cite{CKL}:
\begin{equation}\label{chi}
   \chi = \lim_{\tau \to 0} \Bigl( \lim_{N \to \infty} K( \tau ) \Bigr) \, .
\end{equation}
The meaning of $ \, \chi \, $ is the following:
Let us take a window of the width $ \, \delta E $, $ \, \delta E / \Delta \equiv \bar{n}
\ll N $, in the energy space centered at $E=0$ and calculate the number $ \, n \, $
of levels inside the window at some realization of disorder. The level number variance
is $ \, \Sigma_2( \bar{n} ) = \langle ( n - \bar{n} )^2 \rangle $. The level compressibility
is by definition the limit
\begin{equation}\label{chi-def}
   \chi = \lim_{ \bar{n} \to\infty} \left( \lim_{N\to\infty}
{ \partial \, \Sigma_2 \, ( \bar{n} )
                     \over \partial \, \bar{n} } \right) \, .
\end{equation}
The level compressibility contains an information about the localization transition:
$ \, \chi \, $ ranges from $ \, \chi_{WD} = 0 \, $ for the Wigner-Dyson statistics with
extended wave functions and a strong levels repulsion to $ \,
\chi_{P} = 1 \, $ in the case of localized wave functions and uncorrelated
levels with the Poissonian distribution. The intermediate situation
\[
   0 < \chi_{crit} < 1
\]
is inherent for the critical regime of multifractal wave functions \cite{CKL}.

The exact expression for the level compressibility in the unitary MNS reads \cite{KrMac}:
\RR{
\begin{equation} \label{ChiMNS}
     \chi_{MNS} = \frac{ {\rm Li}_{-\frac{1}{2}} [ 1 - \exp(2 \pi b) ] }
                                                { {\rm Li}_{+\frac{1}{2}} [ 1 - \exp(2 \pi b) ] } \simeq
   \left\{
    \begin{array}{l}
          1/(4 \pi b) \, , \ b \gg 1 \, ; \cr
          1 - \sqrt{2} (\pi b) + [8/\sqrt{3}-2-\sqrt{2}] (\pi b)^{2} \, , \ b \ll 1 \, ;
    \end{array}
   \right.
\end{equation}
}
where $ \, {\rm Li} \, $ is the polylogarithm function
\cite{AbrSt}.  One can see that $ \, \chi_{MNS} \, $ obeys an
inequality $ \, 0 < \chi_{MNS} < 1 $,  at any finite $ \, b  $.

Moreover, the level statistics of MNS and of critical PLBRM are
asymptotically the same in two limits: $ \, b \to 0 \, $ and $ \,
b \to \infty $.

If $ \, b \gg 1 $, the theory of critical PLBRM with $ \, \alpha = 1 \, $ can be rigorously
developed by mapping \cite{MF} onto the nonlinear supersymmetric $\sigma$-model
\cite{Efetov}.  One can show that the level statistics of critical PLBRM approaches the
Wigner-Dyson statistics \cite{KM,ME,KrTs}. In particular, the level compressibility
of PLBRM goes to zero and asymptotically coincides with the compressibility for MNS
\begin{equation} \label{Chilargeb}
        b \gg 1 \Rightarrow \quad
       \chi_{PLBRM}|_{\alpha=1} \simeq \chi_{MNS} \simeq
           \frac{1}{4 \pi \, b} + O(b^{-2}) \ll 1 \, .
\end{equation}
This is because the phase repulsion in MNS is strong at large $ \, b $.  The
phases $ \, \varphi_{i,j} \, $ form an approximately equidistant lattice-like
structure \cite{KM}.

In the opposite case $ \, b \ll 1 \, $, the phase repulsion in MNS is weak and the
phases $ \, \varphi_{i,j} \, $ do not form a regular structure. Disorder in the phase
arrangements at a small distances $ \, | \varphi_i - \varphi_j | \le 1/N \, $ may become
especially important and, therefore, there is no {\it a priori} evident correspondence
between critical PLBRM and MNS at $ \, b\ll 1 $.

Let us consider the $N\rightarrow \infty$ limit of
Eq.(\ref{VarPLperiodic}) at $\alpha=1$. If $ \, b \ll 1 $ the
off-diagonal matrix elements of such a PLBRM are parametrically
small compared to the diagonal ones
\begin{equation}\label{model}
   \alpha = 1, \ b \ll 1: \quad
   \langle \varepsilon_i^2 \rangle = \frac{1}{\beta} \gg
   \langle | V_{ij}|^{2} \rangle \simeq b^2 {\cal F}(i-j) \, , \quad
   {\cal F}(i-j) = \frac{1}{2} \frac{1}{(i-j)^2} \, .
\end{equation}
We will refer to Eq.(\ref{model}) as to the {\it almost diagonal
critical} PLBRMs.  The parameter $ \, \beta \, $ corresponds to
the Dyson symmetry classes: $ \, \beta_{GOE} = 1 $ for the
Gaussian orthogonal ensemble, and $ \ \beta_{GUE} = 2 $ for the
Gaussian unitary ensemble.

This model cannot be mapped onto the nonlinear sigma model as the
mapping is only justified if $ \, b \gg 1 $. At $ \, b \ll 1 $,
the compressibility of PLBRM and MNS are close to the Poissonian
value $ \, \chi_P = 1 $. The leading correction of the order of $
\, O(b) \, $ was derived in Refs.[\onlinecite{ME, Machin}] using
an approximation of two interacting levels first suggested in 
Ref.[\onlinecite{Levitov}]. Surprisingly, disorder in the arrangement 
of MNS phases does not influence $ \, \chi \, $ and the compressibility
for PLBRM and MNS are again asymptotically the same:
\begin{equation} \label{ChiMNSsmallb}
        b \ll 1 \Rightarrow \quad
       \chi_P - \chi_{PLBRM}|_{\alpha=1} \simeq \chi_P - \chi_{MNS}
           \simeq \sqrt{2} \pi \, b + O(b^{2}) \ll 1 \, .
\end{equation}

\subsection{Formulation of the problem}

A natural question arises as to whether the level rigidity of
critical PLBRM and MNS coincide at an arbitrary $ \, b \sim 1 $.
The numerical simulations \cite{KrMac} did not exclude such a
possibility. The main result of the paper is that it is not the
case: the sub-leading corrections of order $ \, O(b^2) $ are
different in those two models. To prove this statement we
analytically calculate {\it the second coefficient of the virial
expansion} \cite{Machin} for the level compressibility for the
critical PLBRM of the unitary symmetry class and compare it with
the exact result Eq.(\ref{ChiMNS}) for MNS.

As the analytical calculation is quite involved we undertook an
extensive numerical investigation of the same problem and found an
excellent agreement with the analytical prediction.

The paper is organized as follows: we briefly discuss the virial
expansion in Section \ref{Virial} and re-derive
Eq.(\ref{ChiMNSsmallb}) as the first virial coefficient in Section
\ref{Virial-1}. The main result of the present paper, namely, {\it
the second virial coefficient} for unitary critical PLBRM, is
calculated in Section \ref{Virial-2}. The analytical result is
confirmed by the direct numerical simulations which are presented
in the Section \ref{NumTst}. We end the paper with a brief
discussion and Conclusions.

\section{The virial expansion}\label{Virial}

The virial expansion (VE) is a method that allows to study spectral statistics
of a disordered system described by a Gaussian ensemble of the Hermitian
$ \, N \times N \, $  ($ N \gg 1 $) almost diagonal random matrices which
have random independent elements \cite{Machin,DOS}:
\[
    \langle H_{i,j} \rangle = 0 \, ; \quad
    \langle H_{i,i}^2 \rangle \gg \langle |H_{i\ne j}|^2 \rangle \, .
\]
It is an {\it expansion in the number of interacting energy
levels}. Unlike the field-theoretical approach, VE starts from the
Poissonian statistics and yields a {\it regular expansion} in
powers of the small parameter controlling the ratio of the
off-diagonal elements to the diagonal ones $ \, \langle | H_{i \ne
j} |^2 \rangle / \langle H_{ii}^2 \rangle \sim b^2 \ll 1 $. The
expansion has been represented by the summation of diagrams which
are generated with the help of the Trotter formula. A rigorous
selection rule has been established for the diagrams, which allows
to account for exact contributions of a given number of resonant
and non-resonant interacting levels. The method offers a
controllable way to find an answer to the question when a weak
interaction of levels can drive the system from localization
toward criticality and delocalization. An example of the spectral
form-factor has been considered in Ref.
 [\onlinecite{Machin}] for a generic dependence of the variance $ \, \langle | H_{i \ne j}
|^2 \rangle \, $ on the difference $ \, i - j $. It has been shown
that a term of the order of $ \, b^{c-1} \, $ is governed by the
interaction of $ \, c \, $ energy levels. VE has been applied to
DOS in Ref. [\onlinecite{DOS}].

VE has been described in detail in Ref. [\onlinecite{Machin}].
Here, we repeat only its basic definitions and final results which
will be applied to the model (\ref{model}). VE deals with the
following correlation function in the time domain:
\begin{equation} \label{K0}
\tilde{K}(N,\tau)=\displaystyle \frac{1}{N}
            \langle\langle \, {\rm Tr} \, e^{ -{\bf i} \, \hat{H} \, \tau / \Delta } \,
                              {\rm Tr} \, e^{  {\bf i} \, \hat{H} \, \tau / \Delta }
            \, \rangle\rangle \equiv \tilde K_0(N,\tau) + b \tilde K_1(N,\tau) +
                                         b^2 \tilde K_2(N,\tau) + \ldots
\end{equation}
For the constant mean density of states $ \, \tilde{K} \, $
coincides with $ \, K $. However, they are different if $\langle
\rho(E)\rangle$ essentially depends on energy $E$. By analogy with
Eq.(\ref{chi}) one can define the quantity
\RR{
\begin{equation}
   \chi_0^{(j)} \equiv \lim_{\tau \to 0} \Bigl( \, \lim_{N \to \infty} \tilde{K}_j (N,\tau) \Bigr) , \ 
   j = 1, 2 \ldots.
\end{equation}
The level compressibility $ \, \chi \, $ can be expressed in terms of virial coefficients 
$ \, \chi_0^{(j)} $ (see Ref. [\onlinecite{SuSy-VE-GUE}] for details):
\begin{equation} \label{K-chi}
   \chi = 1 + \sum_{j = 1}^{\infty} b^{j} \chi^{(j)} , \ 
   \chi^{(j)} \equiv \frac{\chi_0^{(j)}}{\Upsilon_{j+1}} \, .
\end{equation}
Here we have introduced unfolding factors:
\[
\Bigl( \Upsilon_m \Bigr)^{-1} = \sqrt{m} \, .
\]
}
%
%
Each function $ \, \tilde K_i \, $ is governed by the interaction
of the $ \, i+1 \, $ energy levels. The perturbative expansion
(\ref{K0}) is valid if the limit $ \, \lim_{N \to\infty} (\tilde
K_i) $ is finite. This can be secured by a separation of scales:
the level interaction is effectively large at the distances $ \,
|\omega| < \Omega_{int} = b \, \Delta \, $ which are
parametrically smaller than the mean level spacing $\Delta$.
Otherwise, VE fails and one has to take into account an infinite
number of the interacting levels.

\section{Leading correction to Poissonian level compressibility}\label{Virial-1}

The expansion (\ref{K0}) starts with the Poissonian form-factor $ \, K_P $
\[
    \lim_{N \to \infty} \tilde{K}_0 = K_P = 1
\]
reflecting a distribution of uncorrelated diagonal matrix elements.
The functions $ \, \tilde K_i \, $ are given by power series in a large parameter
\begin{equation}\label{Xdef}
   x = \tilde{N} | \tau | b , \quad
   \tilde{N} \equiv \Delta^{-1} \propto N \, .
\end{equation}
The first correction $ \, b \tilde{K}_1 \, $ to the Poissonian spectral statistics in governed by
the interaction of two energy levels. The series for the function $ \, \tilde{K}_1 \, $ in GOE
and GUE reads:
\begin{eqnarray}\label{Ser2Col}
  \tilde{K}_1 & = &  2 \, \sqrt{\pi\beta} \
           \sum_{k=1}^{\infty} (-1)^k \, C^{(2)}_{\beta}(k) \, {\cal R}^{(1)}_N(k) \ x^{2k-1} \, ; \\
  \label{Coef2ColGOE}
  C^{(2)}_{\beta=1}(k) & = &             { (2k-1)!! \over k! (k-1)! } \, ;   \\
  \label{Coef2ColGUE}
  C^{(2)}_{\beta=2}(k) & = &  { 1 \over (k-1)! } \, .
\end{eqnarray}
In Eq.(\ref{Ser2Col}) we introduce the real space sum which
depends on the correlation function $ \, {\cal F} \, $ defined in
Eq.(\ref{model}):
\begin{equation}\label{RSS-1}
     {\cal R}^{(1)}_N (k) \equiv
       \frac{1}{2} {\sum_m}' \Bigl(  {\cal F}(m) \, \Bigr)^k =
       \frac{ \zeta(2k) }{2^{2k}} + O(1/N) \, , \quad
     {\sum_m}' = \sum_{m=-N}^{-1} +\sum_{m=1}^{N} \, ;
\end{equation}
where $ \, \zeta \, $ is the Riemann zeta function \cite{AbrSt}.
The $ \, 1/N$-corrections in (\ref{RSS-1}) yield the dependence of
$ \, \tilde{K}_1 \, $ on a parameter $ \, |\tau| b = x/N $.

To derive the compressibility $ \, \chi_0 $
\[
   \chi_0 \simeq 1 + b \chi_{0}^{(1)} ,
\]
we have to put $ \, \tau = 0 \, $ {\it after} doing the limit $ \,
x \to \infty $:
\begin{equation}
      \chi_0^{(1)} = \lim_{\tau \to 0} \Biggl(
                                \lim_{ x\to \infty }
                                   \Bigl( \tilde K_{1}( x, \, \tau b ) \Bigr)
                                                          \Biggr) \, .
\end{equation}
The small time limit means that we have to neglect all the $ \, 1/N$-corrections in Eq.(\ref{RSS-1}).
It is achieved if one substitutes $ \, {\cal R}^{(1)}(k)  = \lim_{N\to\infty} {\cal R}^{(1)}_N (k) \, $
for $ \, {\cal R}^{(1)}_N (k) $:
\begin{equation} \label{RSS-1-mod}
   {\cal R}^{(1)}(k) = \sum_{m=1}^{\infty} \Bigl(  {\cal F}(m) \, \Bigr)^k \, .
\end{equation}
It is convenient to insert Eq.(\ref{RSS-1-mod}) into the series (\ref{Ser2Col}) and to
sum over $ \, k \, $ prior to the summation over m:
\begin{equation}\label{Res2}
   {\tilde K}_{1}(\tau=0) \simeq - { \sqrt{\beta\pi} }  \sum_{m=1}^{\infty}
     { x \over m^2 } \exp\left( - { x^2 \over 2 m^2 } \right)
     \left\{
        \begin{array}{l}
          I_0 \left( { x^2 \over 2 m^2 } \right) - I_1 \left( { x^2 \over 2 m^2 } \right) \, , \
                   \beta = 1 \, ; \cr
          1 \, , \ \beta = 2 \, .
        \end{array}
     \right.
\end{equation}
Here $ \, I_{0,1} ( \cdots) \, $ are the modified Bessel functions \cite{AbrSt}.  The sum over
$ \, m \, $ converges at $ \, m \sim x \gg 1 \, $ therefore it can be converted to the integral $ \,
\int_0^{\infty} {\rm d} m $. After this integration we find:
\begin{equation}\label{Result2}
   \chi^{(1)}_0 |_{\beta = 1} = -2 \, ;  \quad \chi^{(1)}_0 |_{\beta = 2} = -\pi \, .
\end{equation}

\section{Correction to  level compressibility of order $b^{2}$}\label{Virial-2}

\subsection{The second virial coefficient for the critical PLBRM}

Now we focus on the term of the order $ \, O(b^2) \, $ in
Eq.(\ref{K0}), which is governed by the interaction of the three
energy levels. In the unitary case we will be considering below
the expression for $\tilde{K}_{2}$ reads \cite{Machin}:
\begin{equation}\label{Ser3Col}
   \beta=2: \quad
   {\tilde K}_2 = { 2 \over \sqrt{3} } \! \sum_{k_1, \, k_2, \, k_3=0}^{\infty}
                      (-1)^{k_1+k_2+k_3} \ C^{(3)} (k_1, \, k_2, \, k_3) \,
                      {\cal R}_N (k_1, \, k_2, \, k_3) \ x^{ 2(k_1+k_2+k_3)-2}
\end{equation}
\begin{equation}\label{Coef3ColGUE}
   C^{(3)} = { 2 k_1 k_2 k_3 - k_1 k_2 - k_2 k_3 - k_1 k_3
                                         \over \Gamma( k_1 + k_2 + k_3 - 3/2 ) } \
                               \frac{ \Gamma(k_1-1/2) }{ \Gamma(k_1+1) } \,
                               \frac{ \Gamma(k_2-1/2) }{ \Gamma(k_2+1) } \,
                               \frac{ \Gamma(k_3-1/2) }{ \Gamma(k_3+1) } \, ,
\end{equation}
\begin{equation}\label{RSS-3}
   {\cal R}_N ( \{k_{i}\} ) = { 1 \over 6 } \, {\sum_{m,n}}' \Bigl|_{m \ne n}
       \Biggl[
         \Bigl(  {\cal F}(m) \, \Bigr)^{k_1}
         \Bigl(  {\cal F}(n) \, \Bigr)^{k_2}
         \Bigl(  {\cal F}(|m-n|) \, \Bigr)^{k_3}
       \Biggr] \, .
\end{equation}
The series in r.h.s. of Eq.(\ref{Ser3Col}) is three-dimensional.
It cannot be reduced to a product of one-dimensional series
because of the function $ \, \Gamma^{-1}( k_1 + k_2 +k_3 - 3/2 )
\, $ in the coefficient (\ref{Coef3ColGUE}). We will decouple the
sums over the indices $ \, k_{1,2,3} \, $ using an integral
representation \cite{RizhGr} of the function $ \, \Gamma^{-1}( z )
\, $
\begin{equation} \label{InvGamma}
   { 1 \over \Gamma( z ) } = \frac{1}{2\pi \, \imath } \, \int_{C_1} \,
                                     { \exp ( t ) \over t^z } {\rm d} t \, ;
\end{equation}
\begin{figure}
\unitlength1cm
\begin{picture}(8,8)
   \epsfig{file=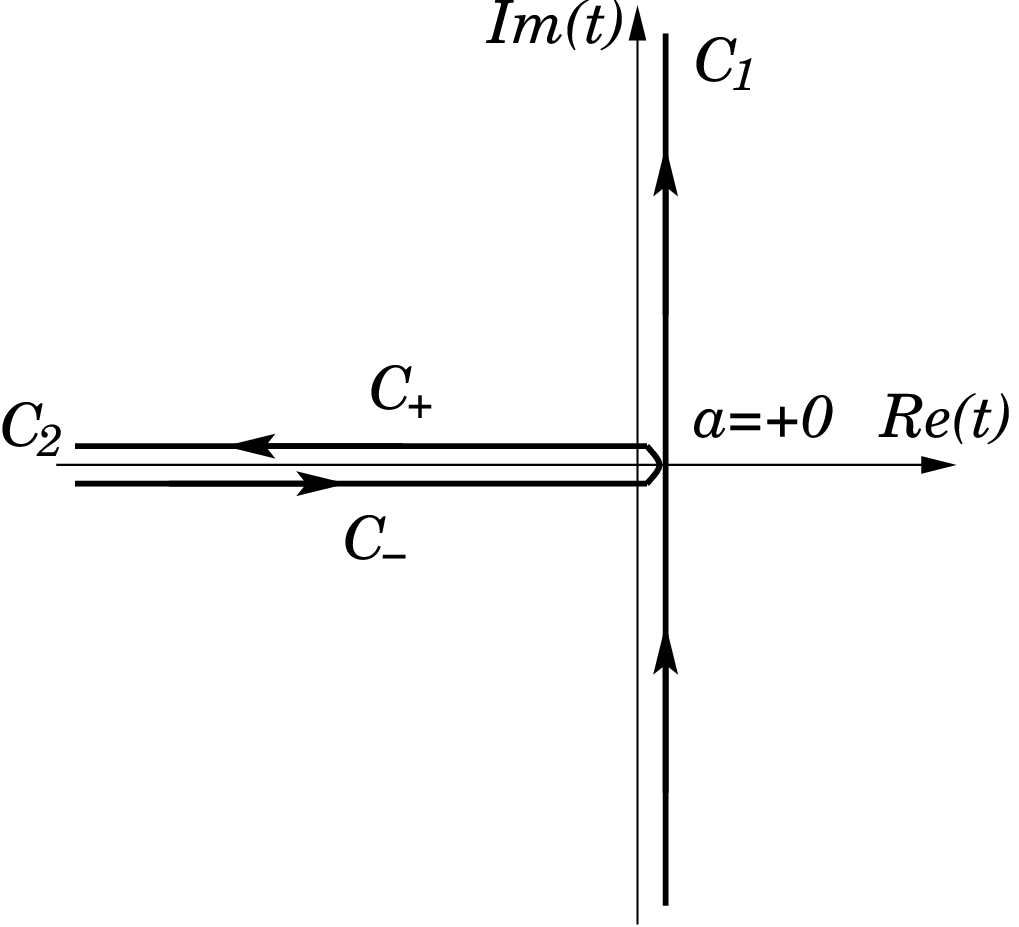,angle=0,width=8cm}
\end{picture}
\vspace{0.5cm}
\caption{
\label{Contours}
The integration contours for the variable $ \, t  $: $ \,  C_1: \{ \, \Re(t) = a >0 \,  , \Im(t) \in
] - \infty; + \infty [ \, \} \, $ and $ \,  C_2 = C_{+} \bigcup C_{-} \, $ where
$ \, C_{-} = \{ \, \Re(t) \in ] - \infty; a ] , \, \Im(t) = -0 \, \} \, $;
$ \, C_{+} =\{ \, \Re(t) \in [ a; - \infty  [ , \, \Im(t) = +0 \, \} $. One can put $ \, a = + 0 $.
}
\end{figure}

\noindent
the integration contour $ \, C_1 \, $ is shown on Fig.\ref{Contours}. Then we change the
order of summations over $ \, k_i \, $ and integration over $ \, t $. The real-space summation
which is implied in the function $ \, {\cal R}_N \, $ has to be done at the last step. We assume
further that the sums over $ \, m \, $ and $ \, n \, $ converge at large values of the summation
variables and transform the sums to a twofold integral (in analogy with the derivation of
$ \, \chi_0^{(1)} $). This assumption is verified below (see Section \ref{Scales}). Using
the identities
\[
   \sum_{k=0}^{\infty} (- y)^k \frac{\Gamma(k-1/2)}{\Gamma(k+1)} =
                   - 2 \sqrt{\pi} \sqrt{1+y} \, ,
\]
\[
   \sum_{k=1}^{\infty} (- y)^k k \, \frac{\Gamma(k-1/2)}{\Gamma(k+1)} =
                  - \sqrt{\pi} \frac{y}{\sqrt{1+y}}
\]
and substituting $ \, \infty \, $ for $ \, N \, $ in the limits of
the real space integrals over $ \, m \, $ and $ \, n \, $ (in
analogy with Eq.(\ref{RSS-1-mod})) we arrive at the following
expression:
\begin{eqnarray}\label{Transf1}
   {\tilde K}_2  (x,\tau=0) & = & { \imath \, x \over 6  } \sqrt{ \frac{\pi}{3} }\,
         \int \!\!\!\int_{-\infty}^{+\infty} \!\! \! {\rm d} m \, {\rm d} n
         \int_{C_1} \!\!{\rm d} t \, \exp (t) \,
             \bar{P}_{|n|}  \bar{P}_{|m|} \Bigl( \bar{P}_{|m-n|} - 3 \, \bar{Q}_{|m-n|} \Bigr) \, ,
\end{eqnarray}
where
\begin{eqnarray} \label{Transf2-1}
       P(y) & = & \frac{y}{\sqrt{1+y}}  , \quad
        \bar{P}_{|j|} \equiv { \sqrt{t} \over x } \, P \left( \frac{ x^2 }{ j^2 t }\right) , \\
       Q(y) & = & \sqrt{1+y}, \quad
       \bar{Q}_{|j|}  \equiv { \sqrt{t} \over x } Q \left( \frac{ x^2 }{ j^2 t }\right) ,
                             \label{Transf2}
\end{eqnarray}
and we have absorbed $ \, 1/\sqrt{2} \, $ into $ \, x \, $
obtaining $ \, 1/2 \, $ as a common prefactor. The integrand in
Eq.(\ref{Transf1}) as the function of $ \, t \, $ has a branching
point $ \, t = 0 $. A branch cut may be drawn along the negative
semi-axis. Since the integrand in Eq.(\ref{Transf1}) is zero if $
\, |t| \to \infty \, $ at $ \, \Re (t) \le a = +0 $ and has poles
neither in upper nor in lower half-plains, we can transform the
integration contour $ \, C_1 \, $ to $ \, C_2 = C_{+} \bigcup
C_{-} \, $ (see Fig.\ref{Contours}) which encloses the branch cut.
Fourier transforming the functions
(\ref{Transf2-1}--\ref{Transf2}) and using a scaled spatial
coordinate
\[
    J = j / x \,  ,
\]
we find the $ \, x$-independent expression for
$ \,  {\tilde K}_2 $:
\begin{eqnarray}\label{Transf3}
   {\tilde K}_2  (\tau=0) & = & { 4 \, \imath \over 3  \sqrt{ 3 \, \pi} } \,
         \lim_{\eta \to +0} \left\{
         \int_{0}^{+\infty} \!\! \! {\rm d} {\cal M}
         \int_{C_2} \!\!{\rm d} t \, \exp (t) \, F_1^2 \Bigl( F_1 - 3 \, F_2 \Bigr)
                                          \right\} \, ; \\ \cr
\label{Fouriers1}
    F_1 \left( {\cal M}, t \right) & = & \int_{\eta}^{+\infty} {\rm d} J \,
           \frac{ \cos( J {\cal M} ) }{ J \, \sqrt{ 1 + J^2 t } } \, , \\
\label{Fouriers2}
    F_2 \left( {\cal M}, t \right) & = & \int_{\eta}^{+\infty} {\rm d} J \,
           \frac{ \cos( J {\cal M} ) }{ J } \sqrt{ 1 + J^2 t } \, .
\end{eqnarray}
We have introduced an infinitesimal positive constant $ \, \eta $, which regularizes
the integrals (\ref{Fouriers1}-\ref{Fouriers2}) at small distances. We
will show that $ \, {\tilde K}_2  (\tau=0) \, $ is finite at $ \, \eta \to +0 \, $ thus
proving that the small distances do not play a role.

Let us separate out real and imaginary parts of parts of $ \, F_{1,2} ( {\cal M}, t \in C_2) \, $
accounting for a branch-cut of $ \, F_{1,2} $ as functions of $ \, t \, $ along the negative semi-axis:
\[
    F_{1,2} |_{ t \in C_2 } = F_{1,2}^{(-)}( M ) +
                   \imath \, {\rm sign}\Bigl( \Im(t) \Bigr) \, F_{1,2}^{(+)}( M ) \, , \
    M \equiv \frac{ {\cal M} }{\sqrt{|t|}} \, , \
    {\rm sign}\Bigl( \Im(t) \Bigr) = \left\{
              \begin{array}{l}
                  1, \mbox{ if } t \in C_{+}   \, , \cr
                  - 1, \mbox{ if } t \in C_{-} \, ;
              \end{array}
                                                          \right.
\]
\begin{eqnarray}\label{Cuts}
    F_{1}^{(+)} & = & \RR{-} \int_{1}^{+\infty} {\rm d} J'
           \frac{ \cos( M \, J' ) }{  J' \, \RR{ \sqrt{ (J')^2 - 1 } } } =
  - \frac{\pi}{2} \int_{ M }^{\infty} {\rm d} y \, J_0 (y) \, , \\
    F_{2}^{(+)} & = & \int_{1}^{+\infty} {\rm d} J'
           \frac{ \cos( M \, J' ) }{ \RR{ J' } } \RR{ \sqrt{ (J')^2 - 1 } } =
           F_{1}^{(+)} ( M ) + \frac{\pi}{2}
          \left( 2 \delta ( M ) - J_1 ( M ) \right) \, ;\\ \cr \cr
    F_{1}^{(-)} & = & \int_{\eta}^{1} {\rm d} J'
           \frac{ \cos( M \, J' ) }{ J' \, \sqrt{ 1 - (J')^2 } } =
  - \frac{\pi}{2} \int_{0}^{ M } {\rm d} y \, H_0 (y) +
         \left[  \RR{\log(2M) + \gamma} -
                  {\rm Ci} \left( \eta \, M \right)\right] \, , \\
    F_{2}^{(-)} & = & \int_{\eta}^{1} {\rm d} J'
           \frac{ \cos( M \, J' ) }{ J' } { \sqrt{ 1 - (J')^2 } } =
           F_{1}^{ \RR{(-)} } ( M ) - \frac{\pi}{2} H_{-1} ( M ) \, .
\end{eqnarray}
Here, $ \, J_{0,1} \, $ are the Bessel functions, $ \, H_{0,-1} \, $ are the Struve functions,
$ \, {\rm Ci} \, $ is the cosine integral function \cite{AbrSt}. We will use the property:
\[
    \lim_{\eta \to +0} \left[ \RR{ \log(2M) + \gamma } -
                  {\rm Ci} \left( \eta \, M \right) + \RR{ \log ( \eta/2 ) } \right] = 0 \, .
\]

One can see that the integral of the real part of $ \, F_1^2 \Bigl( F_1 - 3 \, F_2 \Bigr) \, $
over $ \, t \, $ is zero due to a cancellation of the integrals over $ \, C_{+} \, $ and $ \, C_{-} $.
Thus, we may keep only the imaginary part of $ \, F_1^2 \Bigl( F_1 - 3 \, F_2 \Bigr) \, $ in
Eq.(\ref{Transf3}):
\begin{eqnarray}\label{Transf4}
    F_1^2 \Bigl( F_1 - 3 \, F_2 \Bigr) & \rightarrow &
            \imath \, {\rm sign}\Bigl( \Im(t) \Bigr) \,
                   \Bigl[ \Theta_1( M ) + \Theta_2( M ) \Bigr] \, , \\
             \Theta_1( M ) & \equiv &
                         3 \left( F_{1}^{(+)} \right)^2  F_{2}^{(+)} - \left( F_{1}^{(+)} \right)^3 , \\
             \Theta_2( M ) & \equiv &
                        3 \left( F_{1}^{(-)} \right)^2 \left( F_{1}^{(+)} - F_{2}^{(+)} \right) -
                        6 F_{1}^{(-)} F_{2}^{(-)} F_{1}^{(+)} \, .
\end{eqnarray}
We insert Eq.(\ref{Transf4}) into the expression (\ref{Transf3})
\begin{equation}\label{domain}
   \imath \,
   \int_{0}^{+\infty} \!\! \! {\rm d} \, {\cal M}  F_1^2 \Bigl( F_1 - 3 \, F_2 \Bigr)
   \int_{C_2} \!\!{\rm d} t \, \exp (t) \rightarrow
    2 \,  \int_{0}^{+\infty} \!\! \! {\rm d} M \,
               \Bigl( \Theta_1 + \Theta_2 \Bigr)
                    \int_0^{\infty}  \! {\rm d} t \, \sqrt{t} \, \exp (-t) \, ,
\end{equation}
and integrate over $ \, t \, $ obtaining
\begin{equation}\label{Transf5}
   \chi_0^{(2)} =
         { 4 \over 3  \sqrt{ 3 } } \,
         \lim_{\eta \to +0} \left\{
               \int_{0}^{+\infty} \!\! \! {\rm d} M \,
               \Bigl( \Theta_1( M ) + \Theta_2( M ) \Bigr)
                                         \right\}\, .
 \end{equation}
Let us consider the first integral in r.h.s of Eq.(\ref{Transf5})

\begin{eqnarray}\label{Int1}
    {\cal I}_1 & \equiv & \int_{0}^{+\infty} \!\! \! {\rm d} M \,
             \Theta_1( M ) = \\
                     & = & \left( \frac{\pi}{2} \right)^3
    \int_{0}^{+\infty} \!\! \! {\rm d} M
    \left\{
                3 \left(
                        \int_{ M }^{\infty} {\rm d} y \, J_0 (y)
                   \right)^2
          \Bigl( 2 \delta ( M ) - J_1 ( M ) \Bigr)
               - 2 \left(
                        \int_{ M }^{\infty} {\rm d} y \, J_0 (y)
                   \right)^3
    \right\} \, .
  \nonumber
\end{eqnarray}
We note that the integral Eq.(\ref{Int1}) does not contain the regularizer $ \, \eta $.
The first term with the $\delta$-function function can be immediately integrated using
$ \, \int_{ 0 }^{\infty} {\rm d} y \, J_0 (y) =1 $ [\onlinecite{RizhGr}].
The other two terms can be integrated by parts with the help of the standard integrals
containing the Bessel functions \cite{RizhGr}. The result reads:
\begin{eqnarray}
    {\cal I}_1 =
    \frac{ 3 \, \pi^3 }{4}
    \int_{0}^{\infty} \!\! \! {\rm d} M
    \left( M \, \left[ J_0 \left( M \right) \right]^3 \right)
     = \frac{ \sqrt{3} \, \pi^2 }{2} .
\label{Int1-transf}
\end{eqnarray}

Finally, we have to calculate the second integral in r.h.s of Eq.(\ref{Transf5})
\begin{eqnarray}\label{Int2}
    {\cal I}_2 & \equiv & \int_{0}^{+\infty} \!\! \! {\rm d} M \,
             \Theta_2( M ) = 6 \left( \frac{\pi}{2} \right)^3
              \lim_{\Omega \to + \infty} \Biggl[ \\
\nonumber
                    & \ & \int_{0}^{\Omega} \!\! \! {\rm d} M
    \Biggl\{
                \left( \int_{0}^{ M } {\rm d} y \, H_0 (y)  - \log ( \eta )
                 \right)^2
                \left(  \frac{ J_1 ( M ) }{2} - \delta ( M )
                           + \int_{ M }^{\infty} {\rm d} y \, J_0 (y)
                \right) + \\
\nonumber
                    & \qquad & \qquad +
                   H_{-1} ( M )
                   \left( \int_{0}^{ M } {\rm d} y \, H_0 (y) - \log ( \eta )
                   \right)
                                                \int_{ M }^{\infty} {\rm d} y \, J_0 (y)
    \Biggr\} \, \Biggr] \, .
\end{eqnarray}
Unlike the integral $ \, {\cal I}_1 $, the regularizer of the small distances $ \, \eta \, $
enters the expression for the integral $ \, {\cal I}_2 $. We have also introduced the upper
limit of the integration over $ \, M \, $ before integrating Eq.(\ref{Int2})
by parts. At intermediate stages, the boundary terms of the integration by parts, which
result from the different parts of the integrand in r.h.s. of (\ref{Int2}),
diverge in the limits $ \, \eta \to 0 \, $ and $ \, \Omega \to \infty $. However, the diverging
contributions {\it exactly cancel out} at the end so that the final answer for $ \, {\cal I}_2 \, $
does not depend on $ \, \eta \, $ and is finite in the limit $ \, \Omega \to \infty $. For example,
the coefficient in front of $ \, \log^2 ( \eta ) \, $
\[
  \lim_{\Omega \to + \infty}
    \left\{
    \int_{0}^{\Omega} \!\! \! {\rm d} M
                \left(  \frac{ J_1 ( M ) }{2} - \delta ( M )
                           + \int_{ M }^{\infty} {\rm d} y \, J_0 (y)
                \right)
    \right\}
\]
is zero because
\[
    \int_{ 0 }^{\infty} {\rm d} y \, J_1 (y) = 2 \int_{ 0 }^{\infty} {\rm d} y \, \delta (y) = 1 \, ,
\]
\[
 \lim_{\Omega \to + \infty}
    \left\{
    \int_{0}^{\Omega} \!\! \! {\rm d} M
          \, \int_{ M }^{\infty} {\rm d} y \, J_0 (y)
    \right\} =
 - \lim_{\Omega \to + \infty} \left\{  \Omega
    \left(
          \int_{ \Omega }^{\infty} {\rm d} y \, \frac{J_1 (y)}{y}
    \right)
    \right\} \propto  \lim_{\Omega \to + \infty} \frac{1}{\sqrt{\Omega}} = 0 \, .
\]
The cancellation of $ \, \log( \eta ) $ can be checked in a similar way. We have thus proven
that the regularization of the Fourier images (\ref{Fouriers1},\ref{Fouriers2}) does not
affect the level compressibility.

We skip a lengthy intermediate integration by parts and present only the answer for
$ \, {\cal I}_2 $:
\begin{equation}\label{Int2-transf}
    {\cal I}_2 = 3 \pi^2 \int_{0}^{\infty} \!\! \! {\rm d} M
    \left( H_0 \left( M \right) \,
             J_0 \left( M \right)
     \right) -
                       \frac{ 9 \pi^3}{4} \int_{0}^{\infty} \!\! \! {\rm d} M
    \left( M
              H_0 \left( M \right)^2 \,
              J_0 \left( M \right)
     \right) \, .
\end{equation}
To our best knowledge, the integrals of a combination of the Bessel function and
the Struve function in r.h.s. of Eq.({\ref{Int2-transf}}) are not included in the standard
handbooks. We describe their calculation in Appendices. Here, we give the results:
\begin{eqnarray}\label{OurInts1}
    \int_{0}^{\infty} \!\! \! {\rm d} M
    \left( H_0 \left( M \right) \,
             J_0 \left( M \right)
    \right) & = & \frac{1}{2} \, ; \\
\int_{0}^{\infty} \!\! \! {\rm d} M
    \left( M
              H_0 \left( M \right)^2 \,
              J_0 \left( M \right)
     \right) & = & \frac{2}{\pi} \left( 1 - \frac{1}{\sqrt{3}} \right)  \, ;
                             \label{OurInts2}
\end{eqnarray}
\begin{equation}  \label{Result3}
     \Rightarrow {\cal I}_2 = \frac{3 \pi^2}{2} \left( \sqrt{3} - 2 \right) \, , \quad
            {\cal I}_1 + {\cal I}_2 = \Bigl( 2 - \sqrt{3} \Bigr) \sqrt{3} \, \pi^2 , \quad
            \chi^{(2)}_0 = \Bigl( 2 - \sqrt{3} \Bigr) \frac{ 4 \, \pi^2 }{3} .
\end{equation}
We insert formulae (\ref{Result2},\ref{Result3}) into Eq.(\ref{K-chi}) and obtain the
expression for the level compressibility of the critical unitary PLBRMs:
\RR{
\begin{equation}\label{Chi-Analit}
   \beta = 2: \quad
   \chi = 1 - \sqrt{2} \, ( \pi b ) + \left(  \frac{8}{\sqrt{3}} - 4 \right) \, ( \pi b )^2 +
              O (b^3) \, .
\end{equation}
}

\subsection{ Characteristic spatial scale that governs the compressibility }
\label{Scales}

Let us estimate the characteristic spatial scale that governs the second virial coefficient
$ \, \chi_0^{(2)} $. Firstly, we note that all integrals over $ \, t \, $ and $ \, M \, $
converge at  $ \, |t|_{char} \sim 1 \, $ (see Eqs.(\ref{domain}) and (\ref{Transf5}))
and $ \, M_{char} \sim 1 \, $  (see Eqs.(\ref{Int1-transf}) and (\ref{OurInts1}--\ref{OurInts2})).
Returning to the spatial variable $ \, j = x J $, we may estimate its characteristic scale:
\[
   j_{char} \sim x J_{char} \sim x / {\cal M}_{char}
      \sim x \sqrt{|t|_{char}} / M_{char} \sim x \gg 1 \, .
\]
Therefore, $ \, \chi_0^{(2)} \, $ is governed by {\it the large distances} $ \, m,n,(m-n)
\sim x \gg 1 \, $ (see Eq.(\ref{Transf1})).  We have verified the self-consistency of our
calculation scheme for $ \, \chi_0^{(2)} $, namely, the replacement of the real space sum
by the integral is justified.

The first virial coefficient $ \, \chi_0^{(1)} \, $ is
also governed by the large distances of the order of $ \, x \, $ (see Eqs.(\ref{Res2})).
We may conclude that  {\it the small distances do not contribute to the compressibility}.
One important consequence is that  the level compressibility is not sensitive
to the periodicity of the boundary conditions: If we recalculated
$ \, \chi_{0}^{(1,2)} \, $ using the spatially periodic variance (\ref{VarPLperiodic})
instead of (\ref{model}) we would again arrive at the same results
(\ref{Result2},\ref{Result3}).

\subsection{ Numerical test of the results }\label{NumTst}

\begin{figure}
\unitlength1cm
\begin{picture}(10,10)
   \epsfig{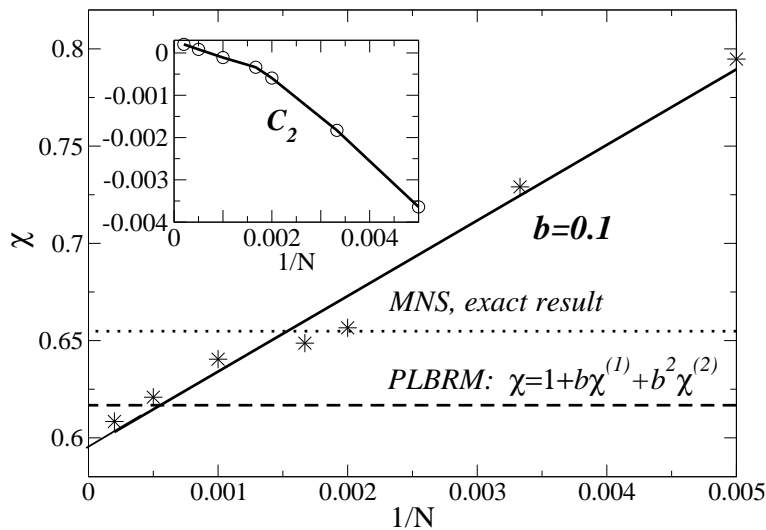}
\end{picture}
\vspace{0.5cm} \caption{ \label{ChiComp} A comparison of the
analytical result (\ref{Chi-Analit}) with the direct numerical
calculation of $ \, \chi \, $ at $ \, b = 0.1 $: stars mark the
result of simulations at the different matrix size; solid line is
an interpolation that yield $ \, \chi|_{N\to\infty} = 0.595 \pm 0.005 $; 
\RR{ dashed horizontal line $ \, \chi_{PLBRM} \simeq 0.617 \, $ is the analytic 
result for $\chi$ with the accuracy up to $ \, O(b^{2}) $ (Eq.(55) for $ \, b=0.1$);}
%
%
dotted line presents the compressibility of unitary MNS. Inset: 
the coefficient $C_{2}$ of the polynomial fitting Eq.(\ref{PolFit})
for~$ \, \Sigma_2 \, $ as a function of~$ \, 1/N $. }
\end{figure}

A comparison of the analytical result (\ref{Chi-Analit}) with the direct numerical
calculation of $ \, \chi \, $ is presented on Fig.\ref{ChiComp}. The data corresponds
to Eq. (\ref{VarPLperiodic}) with $ \, b=0.1 \, $ for the unitary symmetry, $ \,
\beta_{GUE} = 2 $. 

Numerical calculation were done in the range $ \, 200 \le N \le
10\,000 $. The true value of $ \, \chi \, $ is found from the
extrapolation of numerically obtained $ \, \chi(N) \, $ to $ \, N
\to \infty $.

The number of realizations ranges from 30\,000 for the small
matrix size ($N=200$), to 200 for the larger one ($N=10\,000$). The level compressibility
has been obtained from the level number variance $ \, \Sigma_2(\bar{n})=\langle (n-
\bar{n} )^2\rangle $: $ \, \Sigma_2(\bar{n}) \, $ has been calculated in a small
energy window $ \delta E \approx 0.4 $ at the band center \cite{note}. Then, the
$ \, N $-dependent spectral compressibility $ \, \chi(N) \, $ has been obtained
for each value of $ \, N \, $ as a coefficient of the linear term of  a 2nd order polynomial
fit for $ \, \Sigma_2(\bar{n}) $ [\onlinecite{KrMac}]:
\begin{equation}
\label{PolFit}
    \Sigma_2(\bar{n}, N)= C_0 (N) + \chi(N) \bar{n}+ C_2(N) \bar{n}^2 \, .
\end{equation}
The fitting range is $ \, 1 < \bar{n} < 100 $. To increase the accuracy, we have done
both disorder and spectral averaging of the data. We have also checked that a small
change in the window width or in the fitting range practically does not change the results.

For the critical RMT, it is expected \cite{KrMac} that both $ \,
\chi \, $ and $ \, C_2 \, $ have a pronounced ($\propto 1/N$) $ \,
N$-dependence reflecting the finite size effects in $ \,
\Sigma_2(\bar{n},N) $.  The true value of $ \, \chi \, $ is
obtained at the  intersection of the the solid line with the
vertical axis at the point
\begin{equation}
\label{num} \chi|_{N \to \infty} = 0.595 \pm 0.005
\end{equation}
(see Fig.\ref{ChiComp}) which should be compared to the analytical
results for the critical PLBRM and the MNS exact result given by
Eq.(\ref{Chi-Analit}) and Eq.(\ref{ChiMNS}), respectively:
\RR{
\begin{equation}
\label{compare}
   \lim_{N\to\infty}\chi_{PLBRM}(b=0.1) \simeq 0.617 \, ; \quad
   \lim_{N\to\infty}\chi_{MNS}(b=0.1)=0.6548.
\end{equation}
Given that the sign of this omitted correction of order $ \, b^{3}\sim 0.001\, $ 
is negative (by the expected alternation of signs $\chi^{(n)}\propto (-1)^{n}$), 
the difference between the best numerical fit and the analytical approximation 
has a right sign and a right order of magnitude. Thus, the numerical result is 
very close to the analytical one and is clearly different to the MNS exact result.
}
%
%

The true value of  $ \, C_2 \, $ is expected to be zero
\cite{KrMac}. Indeed, the numerically obtained coefficient $ \,
C_2 \, $ goes to zero as $ \, N \, $  increases; see the inset of
Fig.\ref{ChiComp}. This  behavior of  $ \, C_2 \, $ confirms good
quality of our numerics.

\section{Conclusions and Discussion}\label{MNSvsPLBRM} 

As we have already mentioned in the Introduction, the level 
compressibility of the unitary critical
PLBRM and of MNS are asymptotically the same both at $ \, b \gg 1
\, $ [\onlinecite{KM,ME}] and at $b\ll 1$ with the accuracy up to
the terms of order $1/b$ and $b$, respectively. While such a
coincidence is natural at $b\gg 1$, its origin for $b\ll 1$ still
remains unclear.

The main result of the paper, Eq.(\ref{Chi-Analit}), is an
analytical calculation of the level compressibility for the
critical PLBRM ensemble up to the terms of order $b^{2}$ and its
comparison with the corresponding formula for the MNS model of the 
unitary symmetry class.

Our result Eqs.(\ref{Chi-Analit}) shows that the compressibility
in MNS is larger compared to PLBRM
\RR{
\begin{equation}
\label{FinalAnswer}
   \beta = 2: \quad
   \chi |_{\rm MNS} - \chi|_{\rm PLBRM} \simeq 
                       ( 2 - \sqrt{2} )  ( \pi b )^2
\end{equation}
}
in  agreement with the numerical simulations for PLBRM, see
Fig.\ref{ChiComp}. It is also important that the result
Eq.(\ref{Chi-Analit}) is expressible in a simple algebraic form.
The fact that all the intermediate sums and integrals are exactly
doable in terms of elementary functions is not trivial and may
indicate that the PLBRM theory is exactly solvable.

Thus, we conclude that the level compressibility for PLBRM and MNS
are {\it not identical} though very close to each other \cite{KrMac}.  It
is tempting to assume that the coincidence is a consequence of a
certain relation between $\chi(b)$ and $\chi(1/b)$ which holds for
both models, so that the asymptotic coincidence in spectral
statistics for $b\gg 1$ automatically leads to that for $b \ll 1$.
This scenario can also be related with the existence of a
field-theoretical description \cite{Dual} which is {\it dual} to
that of the nonlinear sigma model.

Based on the leading terms in the $b$ and $1/b$ expansion one can
guess a possible form of a relation  between $\chi(b)$ and
$\chi(1/b)$ which can be then checked using the $b^{2}$ and
$1/b^{2}$ terms. From this viewpoint our result
Eq.(\ref{Chi-Analit}) is also a very useful step.

However, maybe the most important conclusion we may draw from the
above consideration is that the {\it virial expansion} method
\cite{Machin} is working and helps to obtain solutions to very
non-trivial problems.

\begin{acknowledgments}

E.C. thanks the FEDER and the Spanish DGI for financial support through
Project No. FIS2004-03117.

\end{acknowledgments}

\section{Erratum: unfolding factor}

\RR{
In the original version of the present paper, we have used an erroneous
relation between $ \, \chi \, $ and $ \, \chi_0 $:
}
\begin{equation} \label{K-chi-WRONG}
 \qquad
 \chi \simeq 1 - \displaystyle { 1 - \chi_0 \over \Upsilon }, \quad \Upsilon =
 \displaystyle \displaystyle\frac{\Delta}{N} \int_{-\infty}^{+\infty}
                        \langle \rho(E) \, \rangle^2 \, {\rm d} E \, , \qquad
 \mbox{WRONG}
\end{equation}
\RR{
which implies constant unfolding factor for $ \, \chi^{(1,2)} $: 
$ \, \Upsilon^{-1} \simeq \sqrt{2} $. Using the recently 
developed supersymmetric version of VE~  \cite{SuSy-VE-GUE,SuSy-VE-GOE}, 
one can show that  unfolding depends on the number of the interacting energy 
levels, see Eq.(\ref{K-chi}); details can be found in Sect. 4.3 of Ref.[\onlinecite{SuSy-VE-GUE}].
Thus, Eq.(\ref{K-chi-WRONG}) is valid only for the leading term of VE. 
}

\RR{
In the present version of the manuscript, we have corrected the subleading terms of
order $ \, O(b^2) \, $ in Eqs.(\ref{Chi-Analit},\ref{compare},\ref{FinalAnswer}).
We emphasize that our conclusions remain valid after correcting unfolding,
namely, the level compressibility in MNS is larger compared to critical PLBRM.
}

\appendix

\subsection*{Appendix: Integrals containing product of Bessel and Struve functions}

In this Appendix we compute the integrals
\begin{equation} \label{IntDef}
       Int_1 = \int_0^{\infty} H_0 (x) J_0 (x) {\rm d} x \, , \quad
       Int_2 = \int_0^{\infty} x H_0^2 (x) J_0 (x) {\rm d} x \, .
\end{equation}

\noindent
{\bf Integral $ \, Int_1 $.} \ Using the integral representation of the Struve function $ \, H_0 \, $
we convert $ \, Int_1 \, $ to the following form
\[
    Int_1 = \frac{2}{\pi} \, \lim_{\alpha \to +0} \,
             \int_{0}^1 \frac{ {\rm d} t }{\sqrt{1-t^2}} \
             \Im \left( \int_0^{\infty} {\rm d} x \, J_0(x)
                                               \exp \Bigl[ ( -\alpha + \imath \, t ) x \Bigr] \right) \, .
\]
The inner integral over $ \, x \, $ is zero \cite{RizhGr}  at $ \, 0 < t < 1 , \, \alpha = 0 $ and
diverges at $ \, t = 1 , \, \alpha = 0 $.
\[
     \int_0^{\infty} {\rm d}x \, J_0 (x) \sin \Bigl( t \, x \Bigr) =
          \left\{
              \begin{array}{l}
                  0 , \mbox{ if } 0 < t < 1 \, , \\
                  \infty, \mbox{ if } t = 1 .
              \end{array}
         \right.
\]
Thus we see that $ \, Int_1 \, $ is determined by the integration over
a small vicinity of the point $ \, t = 1 $. The infinitesimal  parameter $ \, \alpha \, $ has been
introduced to solve an uncertainty $ \, || 0 \times \infty || \, $ with zero coming from the
phase volume at $ \, t \to 1 $: We calculate the integral over $ \, x \, $ keeping the finite $ \, \alpha $:
\begin{eqnarray}\label{I1-transf}
    Int_1 & = & - \frac{1}{\pi} \lim_{\alpha \to +0} \int_0^{1} \frac{ {\rm d} t'}{\sqrt{t'}} \
                     \Im \left\{ \frac{1}{ \sqrt{t' - \imath \, \alpha} } \right\}
          \equiv \frac{1}{\pi} \int_0^{\infty} \frac{ {\rm d} z}{ \sqrt{z(z^2+1)} } \
                      \Re \left\{ \sqrt{- (z+\imath) }\right\} , \ \\
          & \ & t' = 1 - t , \ z = \frac{t'}{\alpha} .
    \nonumber
\end{eqnarray}
The answer (\ref{OurInts1}) results form (\ref{I1-transf}) after the substitution
\[
     - (z+\imath) = \sqrt{z^2+1} \, \exp \left( \imath \,
                                  \left[ \pi + \arctan ( 1/z ) \right] \right) \, .
\]

\noindent
{\bf Integral $ \, Int_2 $.} \ Let us introduce two auxiliary three-fold integrals
\begin{equation}\label{AuxInt1}
    I_2^{(1,2)} = \frac{2}{\pi^2} \int\!\!\!\int_0^1 \frac{ {\rm d}x}{\sqrt{1-x^2}}
                                                                             \frac{ {\rm d}y}{\sqrt{1-y^2}}
                            \int_0^{\infty} {\rm d}q \, q J_0 (q) \cos \Bigl( q [ x \pm y ] \Bigr)
\end{equation}
noting that
\[
    I_2^{(1)} + I_2^{(2)} = \int_0^{\infty} {\rm d}q \, q J_0^3 (q) = \frac{2}{\sqrt{3}\pi} , \quad
    Int_2 =  I_2^{(2)} - I_2^{(1)} \equiv 2 I_2^{(2)} - \frac{2}{\sqrt{3}\pi} .
\]
The idea as how to calculate $ \, I_2^{(2)} \, $ is very similar to the calculation of $ \, I_1 $: we
use the property \cite{RizhGr}
\[
     \int_0^{\infty} {\rm d}q \, q J_0 (q) \cos \Bigl( q [ x - y ] \Bigr) =
          \left\{
              \begin{array}{l}
                  0 , \mbox{ if } -1 < x - y < 1 \, , \\
                  \infty, \mbox{ if } x - y = \pm 1 ,
              \end{array}
         \right .
\]
introduce an infinitesimal regularizing parameter
\begin{equation}\label{AuxInt2}
    I_2^{(2)} = \frac{2}{\pi^2} \lim_{\alpha \to +0} \,
                            \int\!\!\!\int_0^1 \frac{ {\rm d}x}{\sqrt{1-x^2}}
                                                                             \frac{ {\rm d}y}{\sqrt{1-y^2}} \
                      \Re \left\{
               \int_0^{\infty} {\rm d}q \, q J_0 (q) \exp \Bigl( -\alpha q + \imath q [ x - y ] \Bigr)
                             \right\} ,
\end{equation}
integrate over $ \, q \, $ keeping the finite $ \, \alpha $:
\begin{equation}\label{AuxInt3}
    I_2^{(2)} = \frac{2}{\pi^2} \lim_{\alpha \to +0}
                            \int\!\!\int_0^1 \frac{ {\rm d}x}{\sqrt{1-x^2}}
                                                                             \frac{ {\rm d}y}{\sqrt{1-y^2}} \
                      \Re \left\{
     \frac{ - \alpha + \imath [ x - y ] }{ \left( 1+  \Bigl( - \alpha + \imath [ x - y ] \Bigr)^2 \right)^{3/2}}
                             \right\} ,
\end{equation}
and consider only contribution of two small regions $ \, \{ 1 - x \ll 1, y \ll 1 \} \, $ and $ \, \{ x
\ll 1, 1 - y \ll 1 \} $. After lengthy but rather simple algebra we obtain
\[
     I_2^{(2)} = \frac{1}{\pi}
\]
arriving at the answer (\ref{OurInts2}).


\begin{thebibliography}{50}

\bibitem{MF} A.D. Mirlin, Y.V. Fyodorov, F.M. Dittes, J. Quezada, and T.H. Seligman,
                      \pre {\bf 54}, 3221 (1996).

\bibitem{ME} F. Evers and A.D. Mirlin, \prl {\bf 84} 3690 (2000); \prb {\bf 62}, 7920 (2000).

\bibitem{KM} V.E. Kravtsov, K.A. Muttalib, \prl {\bf 79}, 1913 (1997).

\bibitem{KickRotPLRM} B.B. Hu, B.W. Li, J. Liu, Y. Gu, \prl {\bf 82}, 4224 (1999).

\bibitem{AltLev} B.L. Altshuler and L.S. Levitov, Phys. Rep. {\bf 288}, 487 (1997).

\bibitem{Levitov} L.S. Levitov, \prl {\bf 64}, 547 (1990), Annalen der Physik
                             {\bf 8}, 697 (1999).

\bibitem{KrTs} V.E. Kravtsov, A.M. Tsvelik, \prb {\bf 62}, 9888 (2000).

\bibitem{CS} F. Calogero, J. Math. Phys {\bf 10}, 2191 (1969); {\bf 10}, 2197 (1969)
             and {\bf 12}, 419 (1971).
             B. Sutherland, J. Math. Phys {\bf 12}, 246 (1971) and {\bf 12}, 251 (1971).

\bibitem{Garcia} A.M. Garc{\'i}a-Garc{\'i}a and J.J.M. Verbaarshot, \pre {\bf 67},
                 046104 (2003).

\bibitem{MNS} M. Moshe, H. Neuberger and B. Shapiro, \prl {\bf 73}, 1497 (1994).

\bibitem{Gaudin} M. Gaudin, Nuclear Physics {\bf 85}, 545 (1966).

\bibitem{CKL} J. T. Chalker, V. E. Kravtsov and I. V. Lerner, JETP Lett. {\bf 64} (1996), 386.
                       V. E. Kravtsov, Ann. Phys. (Leipzig) {\bf 8} (1999), 621.

\bibitem{KrMac} M.L. Ndwana and V.E. Kravtsov, Journal of Physics A {\bf 36}
                            3639 (2003).

\bibitem{AbrSt} M. Abramowitz, I.A. Stegun, {\it Handbook of mathematical functions with formulas,
                          graphs, and mathematical tables}, Washington, D.C., National Bureau of
                          Standards, 1964.

\bibitem{Efetov} K. Efetov, {\it Supersymmetry in disorder and chaos}, Cambridge, University
                           Press (1997).

\bibitem{Machin} O. Yevtushenko, V.E. Kravtsov, Journ. Phys. A {\bf 36}, 8265 (2003).

\bibitem{DOS} O. Yevtushenko and V.E. Kravtsov,  \pre {\bf 69}, 026104 (2004).

\bibitem{RizhGr} I.S. Gradshtejn, I.M. Ryzhik, A. Jeffrey, D. Zwillinger, {\it Table of
                 integrals, series, and products}, San Diego, CA, Academic Press (2000).

\bibitem{note} For the small matrix sizes the window was slightly increased in order to
                        get sufficient averaged number of the energy levels (around 100) inside the window.


\bibitem{Dual} \RR{
A. Ossipov, V. E. Kravtsov, Phys. Rev. B {\bf 73}, 033105 (2006).
}

\bibitem{SuSy-VE-GUE} \RR{
O. Yevtushenko and A. Ossipov, J. Phys. A: Math. Theor. {\bf 40}, 4691 (2007).
}


\bibitem{SuSy-VE-GOE} \RR{
S. Kronmüller, O.M. Yevtushenko, and E. Cuevas, 
                 J. Phys. A: Math. Theor. {\bf 43}, 075001 (2010).
}


\end{thebibliography}
\end{document}